## Universality in one dimensional orbital wave ordering in spinel and related compounds: an experimental perspective

M. Croft, V. Kiryukhin, Y. Horibe, and S-W. Cheong
Rutgers Center for Emergent Materials and Department of Physics and Astronomy,
Rutgers University, Piscataway, NJ 08854
Brookhaven National Laboratory, Upton, NY 11973

## Abstract

Recent state-of-the-art crystallographic investigations of transition metal spinel compounds have revealed that the d-orbital charge carriers undergo ordering transitions with the formation of local "molecular bonding" units such as dimers in $MgTi_2O_4$, octomers in $CuIr_2S_4$, and heptamers in $AlV_2O_4$. Herein, we provide a unifying scheme involving one-dimensional orbital wave ordering applicable to all of these spinels. The relative phase of the orbitals in the chains is shown to be crucial to the formation of different local units, and thus both the amplitude and phase of the orbital wave play important roles. Examination of Horibe et al.'s [1] structure for $AlV_2O_4$ serves as the vehicle for developing the general behavior for such orbital wave ordering. Ordered $AlV_2O_4$ will be seen to organize into three equivalent chains in 2D Kagome planes coupled so as to form units of three dimer bonds. Three additional equivalent chains manifest a more complex tetramerization with three different charge states and two different bonding schemes. The orbital wave ordering scheme developed is extended to other spinel and related compounds with local triangular transition metal coordination and partial filling of the $t_{2g}$-d orbitals.

## 1. Introduction

The directionality of atomic orbitals, and their bonding, can often play crucial roles in determining the crystal structure in the solid-state. The classic example of this is of course the role of the $sp^3$ orbitals in the covalent bonding in the diamond structures of C, Si and Ge [2]. The d orbitals of transition metals are also very directional; however, in many transition metal compounds like the spinels, the high temperature formation of the basic structure depends only on the valence and atomic size of the transition metal (T) [3]. In some materials the T-d orbital directionality manifests itself through an electronically driven, symmetry lowering, structural transition at a lower temperature. Such transitions are of the general Jahn-Teller type, where T-d orbital ordering occurs within the host structure [4-7]. In this paper, a class of such charge and/or orbital ordering transitions will be examined. Although extensions to a number of structures will be considered, the central vehicle for exploring the role of direct $t_{2g}$ d-orbital overlap in these systems will be the spinel structure [3]. Accordingly, a brief digression on the spinel crystal structure, and transition metal compounds which form in it, will be made.

The cubic spinel $AB_2O_4$ structure (see Figure 1a) consists of tetragonal $AO_4$ and octahedral $BO_6$ structural units. In general, both the A and B sites can contain transition metal atoms. Since the identification in antiquity of the spinel $Fe_3O_4$ (lodestone) as the very first magnetic material, transition metal spinel compounds have played important roles in fundamental solid-state and applied science. On the fundamental side, it is interesting to note that the detailed structure and origin of the low-temperature Verway magnetic/metal-insulator transition in $Fe_3O_4$ have remained unsolved since 1939 [8,9]. It appears that the physics underlying the Verway may be closely related to the concepts discussed in this paper [10,11]. On a more general note, the literature on transition metal spinel compounds as magnetic, battery, metal-insulator, and superconducting materials is too long standing and too voluminous to single out individual references.

We will be concentrating in this work on the transition metals located at the octahedral B sites. For specificity we will use the example of the high temperature cubic phase of $AlV_2O_4$ in discussing the general spinel structure. Figure 1a shows a conventional view of the cubic spinel $AlV_2O_4$ unit cell. The V-O bonds have been indicated and the $VO_6$ moieties have been completed beyond the unit cell. The presence of chains of edge sharing $VO_6$ octahedra extending along the cell-edge to cell-edge , 110-type directions' should be noted. As will be discussed at length below, each V site has three such equivalent chains passing through it.

In Figure 2b a less standard, but hereto relevant, view of a plane perpendicular to the 111-type direction in the spinel $AlV_2O_4$ structure is shown. This view emphasizes that any two of the three intersecting edge-sharing $VO_6$ chains at a V-site, defines a plane with a Kagome topology of V sites (see the blue lines in the figure). A third chain would, of course, extend out of the figure at each site. Interestingly, the Al sites (or tetrahedral A sites of the general structure) lie at points directly above and directly below the centers of the hexagonal holes in the Kagome lattice. In the figures which follow, only the octahedral B sites of the general structure, and the lines connecting them, will be shown. With this general orientation to spinel structure compounds in mind, we will turn to the detailed discussion of the role of $t_{2g}$ overlap in charge/orbital ordering.

Over the past few years, a growing class of transition metal chalcogenide spinel compounds has been shown to exhibit complex and exotic combinations of charge, orbital, and magnetic order [1,10-14]. Progress in these materials has been based squarely on the precise determination of these complex ordered structures. Careful structural measurements on the insulating low-temperature phase of $CuIr_2S_4$ [12], for example, proved to be describable within a complex unit cell involving coupled charge ordering and Ir-Ir dimer formation into intertwined octameric units. In the case of $MgTi_2O_4$ [13], dimer Ti-Ti bonds were found to order into a helical (or chiral) structure.

Croft et al. [14] showed the charge ordered dimers in $CuIr_2S_4$ could be naturally viewed in terms of orbital ordering onto linear chains, accompanied by dimer charge ordering on these chains. Here the direct Ir-5d overlap and band filling in these 1D chains of edge sharing $IrO_6$ octahedra was shown to be consistent with a of $Ir^{3+}$-$Ir^{3+}$dimer/$Ir^{4+}$/$Ir^{4+}$ tetramerization along the chains. Independently, Khomskii et al. [10] showed that the orbital ordering, dimer formation and tetramerized chain formation in $CuIr_2S_4$ could be generalized to explain the ordered state in $MgTi_2O_4$.

At 700 K $AlV_2O_4$ transforms from a high temperature cubic spinel structure into a rhombohedrally-distorted low temperature charge ordered phase. Quite recently, Horibe et al. [1] have elegantly determined the precise structure of the charge ordered phase of $AlV_2O_4$. They interpreted their results within a local picture, where two V sites (V2 and V3) organize themselves into a "molecular heptamer" unit with the third (V1) site being unbonded and supporting a free magnetic moment.

The "heptamer" units involve two "trimers" (triangles of V3 atoms with a dimer bond on each leg of the triangle), joined by a common capping V2 site to form two corner sharing tetrahedral (see a Figure 2b). The V2-V3-V2 bonding assumes the electron contributions of ½-1-½ to form two bonds. The interpretation presented by Horibe et al. [1] appears at this time to be appropriate and correct from a local perspective.

The purpose of this article is to indicate that, like the $CuIr_2S_4$, and $MgTi_2O_4$ examples noted above, it is possible to view the $AlV_2O_4$ charge ordered state in terms of orbital order, bond formation, and charge order on the network of linear chains linking the transition metal sites in the spinel structure. Moreover, it will be seen that the extension of this orbital wave (OW) scheme sheds additional insight into the $CuIr_2S_4$, and $MgTi_2O_4$ systems. Note that the term orbital density wave has not been used, because the density term connotes the preeminence of the amplitude of the wave, whereas both the phase and amplitude are crucial to our discussion. Finally, it is shown that similar ideas can be applied to other non-spinel systems in which $d_{xy}$-type orbital ordering induces structures which can be deconstructed into 1D chain units.

The structural ordering so in these materials is quite complex, and the discussion of them can be commensurately so. Therefore, we will provide a detailed outline of the subsequent sections of this paper. A summary of the system parameters, in the order in which they are discussed in the text, is also provided in Table 1 as a guide to the reader. The model, or approach, we propose here involves the decomposition of these complex structures into superimposed component 1D orbital wave chains. This model will be developed in a series of steps, each of which is motivated by facets in the structure of the compound $AlV_2O_4$. In **section 2**, the detailed crystal and bonding structure of $AlV_2O_4$, proposed by Horibe et al. [1], will be reviewed. In **section 3**, the interpretation of this structure in terms of multiple Kagome planes, each containing multiple 1D orbital wave chains, will be discussed. In **section 3.1**, strongest bonded planes, the $V^{2.5+}$ Kagome planes, containing triangles of dimer-bonded sites (dimer-triangles), along with triangles of non-bonded sites will be discussed. (We refer to these combinations as triangle-dimers, since we believe the dimer is the relevant bond and the grouping is related to the topological lattice structure.) These triangle-dimer planes motivate the introduction of multiple 1D dimerized orbital wave chains, which constitute directional-$d_{xy}$, strong bonding versions of Peierls type chains. In **section 3.2**, the notion of the relative phase function in the orbital wave is introduced and related to the phase factors of the atomic orbitals in a tight binding approximation wave function. The concept of a variation in the orbital occupancy, which scales with the square of the orbital phase factor, is also introduced. The orbital wave in these planes are bond-centered, the electron count per chain site is $n_C=1$, and the total electron count per lattice site (in the plane) is $n_T=2$. Since the site orbital occupancy in the $V^{2.5+}$ Kagome planes is constant, detailed discussion of the orbital occupancy factor is deferred until after the subsequent section. In **section 3.3**, the most complex Kagome planes in the structure, involving the repeat sequence of $V^{2.5+}$-$V^{2+}$-$V^{2.5+}$-$V^{3+}$ sites, are discussed. These planes are shown to be decomposable into 1D $V^{2.5+}$ dimer chains of the sort discussed in section 3.1, along with two equivalent tetramer 1D chains. These 1D tetramer chains motivate an extended discussion of the orbital phase and occupancy functional variations in **section 3.4**. The orbital waves in these tetramer chains are seen to be lattice site-centered, with three sites forming a bonding group, and nodes occurring at adjacent sites. The electron count per tetramer chain site is $n_C$=1/2, and charge order is seen to be present along with the bond-order. In **section 4**, a summary of the elements forced into our model-picture to explain the $AlV_2O_4$ structure is presented (in **section 4.1**), and expanded to other materials in the subsequent subsections. In **sections 4.2 and 4.3**, this model-approach is extended to obtain a deeper understanding of the ordering in the $CuIr_2S_4$, and $MgTi_2O_4$ systems. The tetramer chains in $CuIr_2S_4$ system have hole counts per site of $h_T$=1/2 and per-chain-site counts also of $h_C$=1/2. The bond-centered ordered chains in this compound manifest charge as well as dimer order. $MgTi_2O_4$ also has tetramer ordering, but with an electron count per site of $n_T$=1, a chain count of $n_C$=1/2 per site, and alternating chain-dimer directions. In addition, in this section a natural generalization of the $n_T$=1, $n_C$=1/2 $MgTi_2O_4$ system to a hypothetical $n_T$=2, $n_C$=1 system is proposed, with potential experimental systems where it might be realized being noted [15-19]. In **section 4.4**, additional non-spin systems where orbital wave (OW) ordering might be relevant are discussed. The Kagome planes of edge-sharing $V^{3+}O_6$ octahedra (with $n_T$=2 and $n_C$=1) in the complex $A^{1+}V_6O_{11}$ structure are noted as having an identical structure to the dimer planes in $AlV_2O_4$, with the same triangular combinations of dimers [20,21]. Upon noting that filling the hexagonal holes in the Kagome plane structure yields a hexagonal plane, the orbital ordering in the $n_T$=2, $LiVO_2$ system [22-25] is shown to be entirely consistent with crossing 1D orbital waves with a 3a periodicity, and an OW chain electron count of $n_C$=2/3. Here the generalization to layered materials with hexagonal transition metal planes indicates a much wider potential applications for these OW concepts. In **section 5**, some brief concluding remarks are made.

## 2. $AlV_2O_4$ structure and local bonding

Figure 2a shows a schematic (on a long length scale) of the charge ordered phase of $AlV_2O_4$, based on the structure determined by Horibe et al. [1]. Here, as in all subsequent figures, only the octahedral transition metal sites (V in this case) are shown. The V1, V2, and V3 atoms of Horibe et al. [1] are assumed to be $V^{3+}(d^2)$, $V^{2+}$ ($d^3$) and $V^{2.5+}$ ($d^{2.5}$) respectively. The principal building blocks, within this view of the structure, are the $V^{2.5+}$ ($d^{2.5}$) Kagome planes. In these planes each $V^{2.5+}$ ($d^{2.5}$) atom (blue in the figure) is bonded (at a distance of 2.61 Å) [1] to two other similar atoms by dimer bonds (red) forming the triangle-dimers, noted above. In addition, each $V^{2.5+}$ ($d^{2.5}$) atom is coordinated to two other similar atoms (at a distance of 3.14 Å) [1] through "anti-bonds" (black). In Figure 2b, where an expanded view of a portion of the structure is shown, the adjoining red dimer and black anti-bonding triangles can be seen at the center of the figure. Here the notion that one electron is contributed from each of the dimerized $V^{2.5+}$ ($d^{2.5}$) atoms into the strong dimer bond is assumed.

The $V^{2+}$ ($d^3$) sites (yellow) bridge between bonding triangles in adjacent planes through resonant bonds (green) of length 2.81 Å [1]. Since the $V^{2.5+}$ ($d^{2.5}$) site has 2 electrons involved in strong dimer bonds, there is ½ electron free to participate in bonding to a $V^{2+}$ ($d^3$) site. The $V^{2+}$ ($d^3$) site has six green bonds to $V^{2.5+}$ ($d^{2.5}$) sites. Thus the $V^{2+}$ ($d^3$) atom can also contribute ½ electron to bonding with each of the six $V^{2.5+}$ ($d^{2.5}$) atoms coordinated to it. If there had been one electron per atom the green bonds could have formed dimers. The linear $V^{2.5+}(d^{2.5})$-$V^{2+}(d^3)$-$V^{2.5+}(d^{2.5})$ bonding unit can therefore be thought of as a resonance hybrid (in the chemical sense) of two dimer bonds with the occupancy/weight of the each been ½.

The $V^{3+}$ ($d^2$) sites (black) bridge anti-bonding triangles of $V^{2.5+}$ ($d^{2.5}$) sites in adjacent planes, through non-bonding links (purple) of length 3.30 Å [1]. Here the idea is that, since all of the electrons in the $V^{2.5+}$ ($d^{2.5}$) planes are already involved in bonding, none are left for bonding to the $V^{3+}$ ($d^2$) site. Hence the $V^{3+}(d^2)$ site is left with 2 unbound d-electrons and an isolated S=1 magnetic moment [1].

# 3. AlV$_2$O$_4$ structure and bonding: chain view point

## 3.1 Overview and V$^{2.5+}$ dimer Kagome planes in AlV$_2$O$_4$

The description above was essentially a rephrasing of the arguments of Horibe et al. [1]. Our interpretation, of a multiplicity 1D chains, is locally consistent with that of reference 1, but is globally more easily generalized. Figure 3a shows the $V^{2.5+}$ ($d^{2.5}$) Kagome plane of the bonding and anti-bonding triangles on a long length scale. Viewed on this scale the alternating triangles can be seen to be made up of equivalently structured dimerised chains, labeled $d_{\parallel\text{-}1}$, $d_{\parallel\text{-}2}$, and $d_{\parallel\text{-}3}$. Figure 3b shows the schematic of one of these $d_\parallel$ chains. The ordering an a $d_{xy}$-type orbital onto the chain makes overlap of these $d_\parallel$ orbitals along the chain an important electronic parameter. Since one electron per site goes into each $d_\parallel$ orbital, the $d_\parallel$ 1-D band is half-filled, and susceptible to a Peierls-type dimerization transition [10]. The alternating short and long bond lengths are shown in the top of Figure 3b.

The relative phases of the bonding orbitals, which are lowered in energy by the dimerization transition, are shown in the middle portion of Figure 3b. The similar (++ or --) phase sign of the orbitals in the dimer bonds should be noted, as should be the change of phase between adjacent dimers. This change of relative phase removes the bonding frustration between nearest neighbor atoms in adjacent dimers since the +- phase yields an anti-bonding configuration. This alternating dimer phase yields a formal phase repeat distance of 4a, as illustrated in the figure. The formal phase repeat distance of 4a is a natural consequence of the physically observable Peierls distortion occurring in k-space (where k is electron wave number) at k=π/2a [26]. This 1D distortion will be discussed at greater length below. Here the transition of the states near the Fermi energy, from a 3D structure to $d_{xy}$ orbitals ordered onto 1D chains containing strongly bonded dimers, is analogous to a Jahn-Teller distortion lowering electronic energy. The strong analogy between the Peierls distortion and the Jahn-Teller effect has been invoked by both Hofmann [26] and Khomskii et al. [10].

Considered independently, each of the 1D $d_{\parallel\text{-}1}$, $d_{\parallel\text{-}2}$, and $d_{\parallel\text{-}3}$ chains in the $V^{2.5+}$ ($d^{2.5}$) Kagome planes are similar to the dimerization transition in VO$_2$ [27-28]. However, integration of these chains into a coherent 2D Kagome plane produces the alternating bonding and anti-bonding triangles (see Figure 3a). Here the geometrical pattern of the Kagome lattice naturally lends itself to alternating small and large triangle formation without frustration. It should be emphasized that the dimer bonds in the $V^{2.5+}$ plane are fully occupied (by 2 electrons) and form the strongest bonding units in this structure. The $V^{2.5+}$ dimer planes therefore also form the cornerstone of the resulting three dimensional structure.

In Figure 3b (bottom), we display the dimer centered sinusoidal function which describes the orbital phase variation. (See the section below for an extended discussion.) After Hoffmann [26], the 4a repeat distance (double the distortion distance) of the relative phase is consistent with the unit cell doubling and the requirement that the atomic bonding states adjoining the dimers have the same sign while inter dimer links have differing signs.

## 3.2 Orbital phase p(x) and occupancy variation p$^2$ (x): general observation

At this juncture, it is important to underscore that the sinusoidal site atomic orbital phase functions p(x) used here are styled after the heuristic treatment of Hoffmann [26]. Within the tight binding approximation [29], a lattice's wave functions are formed from a linear combination of atomic orbitals. The relative phase factors in this superposition are determined by the lattice periodicity. For example, the 1D tight binding wave function $\psi_k(x)$ for the $d_\parallel$ band with wave number k would be given by

$$\psi_k(x) \sim \sum_{i=1}^{N} e^{ikX_i} \varphi_{d_\parallel}^i \left( x - X_i \right) \qquad (1)$$

Here the $X_i$ is the position of the i'th lattice site (out of N total sites) and $\varphi_{d_\parallel}^i$ is the ordered $d_\parallel$ orbital at that site. $\varphi_{d_\parallel}^i$ will bear strong resemblance to the atomic $d_{xy}$ orbital, but will be modified by the details of the fully self consistent band structure calculation.

The specific wave functions relevant to the insulating phases of the materials discussed here are determined by the relative phase relations at the distorted k=π/na Brillouin zone boundaries, along the various 1D chain directions. These chains will have distorted supercells of na (where "a" in the example above is the V-V distance). After Hoffmann [26] we choose a real-valued sinusoidal phase combination, p(x), of the ±k combinations of the complex exp(ikx) atomic phase factors. In particular, we will consider the bonding orbital wave function. For specificity, using the tight binding approximation this representation of the bonding state lattice wave function becomes

$$\Psi_{|k|}^b(x) \sim \sum_{i=1}^{N} \sin(kX_i + \theta) \, \varphi_{d_\parallel}^i (x - X_i) \qquad (2)$$

where θ is a phase factor that will fix the nodes of this function for a given type of ordering and the b superscript denotes the bonding orbital. In terms of this example the phase function p(x) is given by

$$p(X_i) = \sin(kX_i + \theta) \qquad \text{or} \qquad p(x) = \sin(kx + \theta) \quad . \qquad \text{(3 a) (3b)}$$

Besides being formally allowable in one dimension, this standing wave, rather than the complex traveling wave, representation is motivated by the localization of the electronic states into local bounding states in the insulating phases in the materials considered here. Of course only the relative spatial variations in the phase have meaning, with the absolute phase oscillating in time similar to a standing wave on a string.

The orbital phase function, p(x), facilitates visitation of the orbital wave ordering pattern however, a cautionary note regarding it is warranted. In the dimerization case, for example, the physical lattice and bonding property periodicity change is a→2a (or na in general). On the other hand the periodicity of the dimerized phase function, p(x), is formally 4a (or 2na in general). Thus while the utility in visualization motivates our use of the formal (unphysical) 2na period of p(x), the shorter na period of the physical observables should always be kept in mind for the systems discussed here.

The orbital phase function, p(x), facilitates visitation of the orbital wave ordering pattern however, a cautionary note regarding it is warranted. In the dimerization case, for example, the physical lattice and bonding property periodicity change is a→2a (or na in general). On the other hand the periodicity of the dimerized phase function, p(x), is formally 4a (or 2na in general). Thus while utility in visualization motivates our discussing the formal (unphysical) 2na period of p(x), the shorter na period of the physical observables should always be kept in mind for the systems discussed here.

The real sinusoidal orbital phase variation from site to site allows for the possibility of a modulation in the probability amplitude (or orbital occupancy) at the various sites. The relative orbital occupancy at each site will vary as the square of the orbital phase function $p^2(x)$, namely $n_{d\parallel} = \alpha p^2(x)$, where $\alpha$ is a proportionality constant. Here, for the purpose of clarity of presentation, we will adjust this proportionality constant to yield the average site occupancies consistent with the formal chemical valence in the various compounds. Appropriate electronic structure calculations are required to yield the actual site orbital occupancies.

In the case of a simple dimerized chain, as in Figure 3b, the phase function p(x) is bond centered, assumes the values of ±0.707 at successive lattice sites, and $p^2(x)=0.5$ at all sites. Thus, in this case, the proportionality constant is $\alpha=2$ yielding the constant $n_{d\parallel} = 1$ at all sites. In some of the cases considered below, the orbital phase function will have a non-constant modulus, or even a node, at various lattice sites, causing a strong modulation in the orbital occupation. The notion of the orbital occupancy variation coupling to the orbital phase variation will be returned to below, after introducing an experimental case where the orbital occupancy variation is relevant.

### 3.3 Complex $V^{2.5+}\ V^{2+}\ V^{2.5+}\ V^3$ Kagome planes in $AlV_2O_4$

Planar cross sections of the transition metal sites in the spinel structure involve multiple Kagome planes. In Figure 4a, a Kagome plane that contains all of the types of V sites is shown. In the horizontal direction in Figure 4a, dimer $V^{2.5+}$ chains, of the type shown in Figure 3, can be seen. In addition, however, two new, longer period and more complicated chains d'$_{\parallel-1}$, and d'$_{\parallel-2}$ should be noted. (A third identical, d'$_{\parallel-3}$, chain would lie in another Kagome plane.) The repeat pattern in these new chains is illustrated in Figure 4b. The valence state sequence in this chain is $V^{2.5+}\ V^{2+}\ V^{2.5+}\ V^{3+}$ with the repeat distance for this tetramer being 4a. The bond sequence in this tetramer is non-bonding/resonance-bonding/resonance bonding/non-bonding again with a repeat distance of 4a. These 4a sequences correspond to the physical observables for the chains.

The formal relative phase relation of the $d_\parallel$ orbitals in this $V^{2.5+}\ V^{2+}\ V^{2.5+}\ V^{3+}$ chain is shown in the middle frame of Figure 4b. The linear-trimer sequence $V^{2.5+}\ V^{2+}\ V^{2.5+}$ in the chain involves two resonance bonds, and of the V sites must have the same phase (sign) so that bonding can occur. The $V^{2+}$ site must, however, be free of bonding.

Therefore, the phase of the $V^{2.5+}$ $V^{2+}$ $V^{2.5+}$ linear-trimers on either side of the $V^{2+}$ site must be of opposite sign. In this way the orbital at the $V^{2+}$ site is frustrated, and must at an interface of + and- phase along the chain. Neither the+ or - phases can be stable at this site and this is indicated in the figure by a box at the site containing both orbital phase states. The formal repeat distance for the orbital phase is 8a and is again twice that of the physically observable atom or bond type repeat distance. Again, the alternation of the orbital phase on either side of the $V^{2+}$ site that is crucial in leaving this site in a quasi free d orbital state with a free S=1 magnetic moment [1].

As in the dimerization discussed in the previous section, it is possible to construct a sinusoidal orbital phase relation appropriate for the a$\Rightarrow$4a unit cell increase. This orbital phase will have a formal repeat distance of 8a, and is shown in Figure 4b just below the orbital sketch. It is crucial to note 4 points: first, that the orbital phase must be precisely the same at adjacent $V^{2.5+}$ sites adjoining the $V^{2+}$ site; second, that the orbital phase must be opposite in sign, but of the same magnitude, at sites adjacent to the $V^{3+}$ sites; third, and consistent with the previous point, the $V^{3+}$ sites must be positioned at nodes (or zeros) of the orbital phase function; and finally, that the $V^{2+}$ sites must lie at extrema of the orbital phase variation. The sinusoidal orbital phase variation with all of these points is shown in Figure 4b. The nodal positioning of the $V^{3+}$ sites is what leaves theses sites with decoupled fluctuating orbital and magnetic moment degrees of freedom [1].

### 3.4 Intra chain orbital occupancy variation: general observation

The above experimental example clearly involves substantial variations in orbital occupancy along the chain directions. The energy driving orbital ordering and chain formation in these materials is the direct overlap and bonding between $d_\parallel$ orbitals ordered statically onto the chain [14]. In the metallic phase of these materials a superposition of the $d_{xy}$–type orbitals, with equal weights along all chain directions, is involved [see footnote 30]. It is the relative orbital phase between adjacent chain sites that determines the strength of the bond between these sites. Since the combination of ++ and -- adjacent site phases yields the same bonding interaction, the coupling between the orbital phase variation function and the variation of bonding energy would be expected to be quadratic, to leading order, in the tight binding phase, p(x), noted above. The total energy gain (cost) of a bonding (anti-bonding) orbital scales with the electron count or occupation of that orbital in the chain, again underscoring the quadratic coupling between the orbital phase function, p(x), and the orbital occupancy, i.e. $n(d_\parallel)= \alpha\ p^2(x)$. This relation simply reflects the idea that there should be charge (or orbital weight) transfer to atoms involved in strong bonding, and away from those involved in anti- or non-bonding. Again, the values of $\alpha$ used at various portions of this text are adjusted to yield the formal average electron (hole) count per site from the phase function in the unnormalized tight binding wave functions. Of course the real physical electron counts require detailed electronic structure calculations. Nevertheless the relative orbital occupancy counts quoted here should be representative of the underlying physics.

In the bottom frame of Figure 4b, we plot $p^2(x)=n(d_\parallel)/\alpha$ which represents the expected relative variation of the $d_\parallel$ orbital occupancy. Several points should be noted in this plot. First, $n(d_\parallel)$ must equal 0 at the nodal point of p(x), which is at the position of the free orbital $V^{3+}(d^2)$ site. Thus, while there are d-electrons at this site, they are not involved in or coupled to the orbital ordered chain. In contrast to the bond-centered example considered above, the orbital phase function is atom centered in this class of chains. Second $p^2(x)$ peaks at 1.0 at the $V^{2+}(d^3)$ site, where three d electrons are shared between three equivalent chain directions, so that there is one electron per $d_\parallel$ orbital. Thus the proportionality parameter $\alpha = 1$ is appropriate for this class of chains.

Third, the value of $p^2(x)$ is ½ at the at the $V^{2.5+}(d^{2.5})$ site. This is, of course, in exact agreement with the arguments above, in which the $V^{2.5+}$ site had two d electrons involved in in-plane dimer bonds and ½ electron left for a $d_\parallel$ orbital pointing toward the $V^{2+}$ $d^3$ site. Thus, within the stated assumptions, it would appear that the observed 4a bond and orbital occupation variations are highly consistent with the simple orbital wave scheme proposed here.

## 4. Summary and extension to other systems
### 4.1 Spinel systems

Our results indicate that the complex valence state and bonding state structure of $AlV_2O_4$ can be interpreted within a simple model in terms of the orbital wave ordering onto the 1D chain network of the spinel structure. The periodic bond variations along the chains are consistent with the variations in orbital phase along the chain, and the charge variations follow from a quadratic coupling to the orbital phase variation. With the apparent utility of this scheme in this complex material, it is useful to go back and re-examine the discussion of the experimental systems which gave rise to the notion of 1D chain decomposition in the spinel structure[10-14].

### 4.2 CuIr$_2$S$_4$ revisited

In Figure 5 (top) we present a schematic representation of the 1D tetramerized chains of $Ir^{4+}$-$Ir^{4+}$ dimer/$Ir^{3+}$/$Ir^{3+}$ in $CuIr_2S_4$ [10-14]. The first-step model explanations of this structure invoked, for simplicity, pure $Ir^{3+}$ and $Ir^{4+}$ states [14,10]. The authors of these models clearly recognized the ephemeral nature of pure valence states in such solids however assuming integral valence states provided simplicity to the discussions and made contact with previous experimental interpretations [12]. One drawback of this model was that the pure $Ir^{3+}$-$Ir^{3+}$ pairs of sites, with their entirely-filled-$t_{2d}$-orbitals, suggested that the dimer chains had, in some sense, missing links.

In the Figure 5 (middle), an 8a orbital phase variation, of the same type proposed for the 4a orbital occupancy variation in $AlV_2O_4$, is shown. There is, however, one absolutely crucial difference: in this case of $CuIr_2S_4$, the ±1 extrema of the orbital phase function occur in the center of the $Ir^{4+}$-$Ir^{4+}$ dimer bonds, and the nodes occur **between** the $Ir^{3+}/Ir^{3+}$ pairs. (Recall for $AlV_2O_4$, a node at an atomic site was required.) Thus, the orbital phase function is bond centered in $CuIr_2S_4$. A second difference for $CuIr_2S_4$ is that the **hole count** in the $d_\parallel$ (or $t_{2g}$ band) is, on average, ½ per site. This means that the occupancy oscillation will seek to maximize the <u>hole count</u> in the excited anti-bonding states at the bonding "$Ir^{4+}$-$Ir^{4+}$"dimer sites, and/or the electron count at the nonbonding "$Ir^{3+}$-$Ir^{3+}$" pairs of sites.

Invoking the same quadratic coupling of the orbital occupancy (in this case the hole occupancy) to the orbital phase function, one finds the values of $[p^2(x)=n(d_\parallel)/\alpha]$ at the $Ir^{4+}$ site to be a=0.83 (labeled A in the figure), and at the $Ir^{3+}$ to be b=0.14 (labeled B). Here it is understood that $n(d_\parallel)$ represents the hole (rather than electron) variation. With the assumed average hole count per site of ½, one finds the proportionality constant $\alpha = 1.02$. This yields the values of $n(d_\parallel)$ at the A site of 0.86 and at the B site of 0.14. Again, these values should be considered as nominal, with electronic structure calculations being required for quantitative estimates. Thus it would appear that the notions developed to interpret the structure in $AlV_2O_4$ can be carried over to enhance the picture of the outwardly quite different physics for $CuIr_2S_4$.

### 4.3 $MgTi_2O_4$ revisited and a generalization noted

One of the more striking assumptions in the discussion of $AlV_2O_4$ is that if the d-electron count on a V site is n, and a portion of the count $n_1$ is involved in chain bond formation, then the remaining $n-n_1$ electrons can separately be involved in bonding on other chains. For example, the $V^{2.5+}$ $d^{2.5}$ site has 2 electrons in separate dimer bonds, and the remaining ½ electron is involved in a resonance bond. With this in mind we can reconsider the ordering in $MgTi_2O_4$ [13].

In Figure 6a, a portion of the ordered structure of $MgTi_2O_4$ [13] is shown. The tetramerized chain 1 has a dimer bonded pair $A_1$-$A_1$, followed by a pair of $B_1$ sites which do not possess a bond along the chain (the next $A_1$-$A_1$ dimer is shown for clarity). Each of the $B_1$ sites, on the other hand, are involved in dimers along other chain directions (e.g. chains 2 and 3). The condition that all sites are dimerized (in some direction) is required by the electron count of one electron per site ($n_T=1$) as in $VO_2$ [27-28]. Thus the two B sites are A sites in other chain directions. Again, thinking in terms of integral chain site occupancy this first step model creates the impression that the non-bonded sites are missing links in the dimer chain.

Returning to the schematic shown in Figure 5, the reader should now consider the A and B sites as those of Ti in $MgTi_2O_4$ along the "1" chain in Figure 6a. The $A_1$ sites are dimerised with high orbital occupancy [$n(d_\parallel)$ at $A_1$ ~0.86], and the next two $B_1$ sites are non-bonding with low occupancy [$n(d_\parallel)$ at $B_1$ ~0.14]. Again, the actual numbers are only for the sake of argument. This analogy underscores the fact that the orbital phase function in $MgTi_2O_4$ is again bond-centered. The low occupancy $B_1$ sites, therefore, have almost a full electron (1-0.14=0.86) left to act as "A sites" [see ($A_2$) and ($A_3$) in Figure 6a] on a crossing chain. This approach invokes the notion (developed to understand the $AlV_2O_4$ structure) that electrons, or fractions thereof, not involved in bonding on one chain can bond on another chain. In this case however, it is approximately a single electron that is being partitioned. The orbital state at each Ti site would, in this view, be a superposition of a majority and minority static polarization direction/occupancy, acting as an A site and a B site in crossing chains. This view of $MgTi_2O_4$ has a number of advantages. It unifies the explanations of all three of these systems into a more consistent framework. As in the $CuIr_2S_4$ case, it also makes the continuity of the chains clearer, as well as allowing for expected departures from integral valences/occupancies.

It is instructive to consider the generalization of the dimer bond-ordering in $MgTi_2O_4$ that is shown in Figure 6b. In Figure 6b, additional dimer bonds have been added at the positions of the violet arrows. Doing so generates a new structure in which the electron count at every site is now $n_T=2$. In this $n_T=2$ structure, there are two non-collinear bonds at each site, directed along crossing dimerized chains. The orbital phase period along a chain would now be 4a, and the orbital occupancy period 2a come as in the dimerized chain case.

The overall crystal distortion for this $n_T=2$ case would be tetragonal (as in the in the $n_T=1$, $MgTi_2O_4$ case), as illustrated in Figure 6c. The spiraling of the dimer bonds along the crystal tetragonal c-axis is shown in Figure 6c (top), and the deletion of the bonds indicated by the arrows (along with a length change) would recover the $n_T=1$, $MgTi_2O_4$ structure. The dotted line in Figure 6c (top) indicates one of the 1D-dimerized chain directions. In Figure 6c (middle), a view of the spirals looking down the c-axis is shown for the $n_T=2$ case, and the two equivalent directions of the tetragonal structure should be noted. Similarly in Figure 6c (bottom), the view of the spirals looking down the c-axis is shown for the $n_T=1$, $MgTi_2O_4$ structure again with two equivalent basal plane directions typical of the tetragonal structure (albeit with the obvious ½ factor of bond depletion).

There are transition metal spinel compounds which have $n_T=2$ at the B-lattice sites. Examples are the $AV_2O_4$ (A=Mg, Zn and Cd) spinel compounds [15-19]. These examples do manifest structural transitions at which the magnetic susceptibility decreases as in the systems discussed above. For these $AV_2O_4$ compounds the transitions are indeed cubic to tetragonal and orbital-order/spin-singlets have been invoked in explanations [15-19]. At present, however, there is no evidence for the presence of the $n_T=2$ structure postulated above in these compounds and alternate orbital orderings have been proposed [15-19]. Indeed there is substantial uncertainty as to the detailed structure in the these compounds as illustrated by the reported absence of both the long range structural and magnetic order in single crystal $ZnV_2O_4$ [19]. Thus while there is no definitive realization for this specific $n_T=2$ structure there are multiple candidates where similar 1D orbital order considerations may be relevant.

In the $AV_2O_4$ compounds, reduced magnetic moment ordering at lower temperatures also occurs, and deserves additional comment here [15-19]. For each of the dimer bonds discussed above, there is an excited triplet magnetic state. In the orbital ordered state in such materials, direct, anisotropic exchange interactions along the chains could be anticipated and an exchange splitting of the triplet state introduced. The exchange split excited state could then admix with the singlet, producing a reduced-moment spin density wave on top of the orbital wave.

Summarizing we have presented a simple scheme for unifying the complex orbital, bonding, and charge ordering phenomena in the three spinel systems where precision structural determinations have been made. The scheme presents a blueprint for detailed electronic structure calculations to build upon. Additional experimental or theoretical work is clearly called for to challenge the assumptions of the schematic framework presented here.

## 4.4 Non-spinel systems with 2D layers

It is important to note that the ideas discussed above should also be applicable to non-spinel systems. The first class of materials are $AV_6O_{11}$ with A=K, Na, Sr and Pb [20,21]. These complex materials possess Kagome planes of $V^{3+}$ $d^2$ sites, which form triangle-dimers with precisely the same proposed structure shown in Figure 3a [20,21]. The model presented here for the Kagome triangle-dimer planes (see Figure 3a and related text) should therefore be directly transferable to the Kagome planes of these systems. Indeed, the lack of the additional ½ electron bonding out of these planes, as in the $V^{2.5+}$ Kagome triangle-dimer planes of the $AlV_2O_4$ case, make the 1D dimer chain groupings into triangle-dimer groupings in the $AV_6O_{11}$ Kagome planes a clearer, simpler example of the phenomenon.

The second example, $LiVO_2$, involves generalizing this coupled chain-based scheme to the important class of layered transition metal chalcogenide compounds. This material has planes of edge-sharing $VO_6$ octahedra [21-25]. The V sites form hexagonal planes of $V^{3+}$ $d^2$ atoms, with every in-plane V-V connection being through a ligand octahedral edge. Moreover, the $LiVO_2$ system has been shown to transform into a low temperature nonmagnetic state, interpreted by various authors in terms of V-trimers [24,25]. Again we refer to these combinations as triangle-dimers, since we believe the dimer is the relevant bond, and the grouping is related to the weaker interactions in the topological lattice structure [16,24,25]. Finally, the fact that the hexagonal planes of this system can be obtained from the Kagome lattice by filling its hexagonal holes with a central sites should be noted. These points strongly suggest that an explanation, involving $d_{xy}$-type ordering similar to that used in the Kagome planes noted above, should be possible.

Pen et al. [24] have in fact proposed an orbital ordering for the $LiVO_2$ system which matches the timer grouping (of three dimer V-V bonds) in this hexagonal system and a 3a superstructure, has been observed experimentally [25]. Pen et al [24] however only alluded to a potential chain interpretation for this material, as did Khomslii et al [10]. In Figure 7, a rendering of the ordering proposed by Pen et al [25] is shown, along with our proposal for the orbital phase and occupancy variation for this material. The one type of chain in this structure is shown with the understanding that two identical chains, oriented at 60 ° and 120° degrees, would be required to fill the 2D structure (as in Figure 3a). The chain orbital phase repeat unit is +, +, node, -, -, node with a formal phase repeat distance of 6a (where a is the hexagonal inter-atomic distance). Interestingly in this case, where the physical super cell period is odd (3a), one has an intermediate combination of an atom- and a bond- centered phase function oscillation. The orbital phase function is shown in Figure 7 bottom, along with the orbital occupancy variation. The nodes of the orbital phase function occur at the zero-occupancy sites along the chain. These nodal sites still possess two d-electrons, which are ordered onto other chain directions. The nodal V-sites and the 6a phase period (in contrast to the Kagome triangle-dimer plane case) are mandated by the higher packing density of the triangle-dimer units in the full hexagonal planes. The alternating + and − extrema of the orital phase variations are at the center of the dimer bonds, with the occupancy of all dimers being equal, but with an intervening nodal occupancy site along a given chain.

Although the total formal d-occupancy per V site is two in this system, the orbital occupancy along the chain direction is 2 per 3 sites or 2/3. The structural periodicity change along the chain is a $\Rightarrow$3a, consistent with band gap formation at 1/3 and 2/3 band filling levels, the former of which is operative here.

# 5. Conclusion

A promising leitmotif linking together complex structural transitions in a wide class of transition metal compounds had been discussed in this paper. This class of materials involve: partially filled $t_{2g}$ orbitals; octahedral chalcogenide coordination; and formally identifiable edge sharing octahedral chains (indeed, multiplicities thereof). The important interaction underlying these electronic transitions is the direct overlap of $d_{xy}$-type orbitals across these shared edges. This interaction drives orbital ordering onto the chains. The modulation of the orbital phase and orbital occupancy along the chain direction is determined by the chain orbital filling factor. A summary of the various compounds and their properties discussed in this paper is presented in Table 1. The detailed integration of multiple chains into a 2D or 3D structure depends on their periodicity and the topology of the structure.

The proposed extension of the notion of multiple coexisting 1D orbital waves to layered materials with hexagonal planes suggests consideration of potential orbital wave affects in a multitude of materials. Our group has, for example, recently studied the layered hexagonal planar $Na_xCoO_2$: y $H_2O$ ($1.0 \geq x \geq 0.3$) system, where the formal $t_{2g}$ hole count h~$(1-x)$ [33]. The proximity of multiple Na sites to the $CoO_6$ planes would presumably pin local orbital orientations in the anhydrous (y=0) compounds of this system. Our work has supported the notion that that the dominant effect of hydration, at the important x~0.3 composition, is to render the $CoO_6$ planes more homogeneous due to the spatial displacement of the locally perturbing Na Coulomb potential [33]. It is tempting to consider the possibility that orbital wave fluctuations could play a role in the onset of the non-conventional superconductivity in the hydrated material.

We have been careful to restrict consideration in this paper to systems where orbital wave ordering is traceable to direct $t_{2g}$ overlap between edge-shared octahedral sites. The resulting orbital wave orderings can be thought of as driven by a generalized Jahn-Teller electronic state lowering [10,26]. Jahn-Teller orbital ordering phenomena do exist in other transition metal systems (absent direct $t_{2g}$ overlap). Studies based upon possible similarities between linear ordered morphologies observed in such systems to those described here are worthy of consideration in future work.

**Acknowledgements**

The author would like to thank R. Hoffmann for valuable and timely concept clarification. The authors would also like to thank E. Andrei, G. Kotliar, and P. Chandra for helpful conversations. We also wish to thank William Mayo, of the Rutgers Materials Science and Engineering Department, for valuable discussions and aid in the use of Crystal Impact's Endeavour version 1.2_ software for structure examination. This work was supported in part by the National Science Foundation under grants (DMR-0405682 and DMR-0093143).

| Compound, config. | structure | $n_T$ | chain period | $n_c$ | chain occupancy/ valence sequence | centering | comment |
|---|---|---|---|---|---|---|---|
| $AlV_2O_4$ $V^{2.5+}$ ($d^{2.5}$) | spinel | 2* 2a-plane | 2a | 1 | 1,1 $V^{2.5+}$,$V^{2.5+}$ | bond | 2 dimers per site $V^{2.5+}$ planes |
| | | 2* 4a-plane | 4a | 0.5 | 0.5,1,.0.5,0 $V^{2.5+}$,$V^{2+}$,$V^{2.5+}$,$V^{3+}$ | atom | 1 $V^{2.5+}$ free orbital 2 resonance bonds |
| $CuIr_2S_4$ $Ir^{3.5+}$ ($h^{0.5}$) | spinel | 0.5 (h) | 4a | 0.5 (h) | 0.86,0.86,.14,.14 $Ir^{3.14+}$,$Ir^{3.14+}$,$Ir^{3.86+}$,$Ir^{3.86+}$ | bond | dimer, charge order |
| $MgTi_2O_4$ $Ti^{3+}$ ($d^1$) | spinel | 1 | 4a | 0.5 | 0.86,0.86,.14,.14 $Ti^{3+}$, $Ti^{3+}$, $Ti^{3+}$, $Ti^{3+}$ | bond | 1 dimer per site, A & B chains |
| Proposed ($d^2$) | spinel | 2 | 4a | 1 | 0.86,0.86,.14,.14 constant valence | bond | 2 dimers per site, A & B chains |
| $A^{+1}V_6O_{11}$ (Kag. planes) $V^{3+}$ ($d^2$) | hexagonal p6$_3$/mmc | 2* | 2a | 1 | 1,1 $V^{3+}$,$V^{3+}$ | bond | Just Kagome planes of complex struct. |
| $LiVO_2$ $V^{3+}$ ($d^2$) | layered hexagonal | 2 | 3a | 2/3 | 1,1,0 $V^{3+}$,$V^{3+}$,$V^{3+}$ | atom/ bond | 2 dimers per site |

Table 1.  A table of the properties of the compounds, in the order discussed in the text.  The average transition metal valence and d electron (h=hole) count over all lattice sites is given under the chemical formula.  $n_T$ is the total d-orbital count per site in the compound, or specific planes in the compound, and the (h) for $CuIr_2S_4$ denotes the hole count. The * indicates that specific planes in the compound are referred to.  The "chain's" column refers to the superstructure periodicity where "a" is the inter-transition metal distance in the chain.  $n_C$ is the average d-orbital count per site along a given chain type [again (h) denoted rather than electron count].  The "chain occupancy/valence sequence" column refers to the chain occupancy variation and valence variation along the given chain.  The "centering" refers to whether the orbital wave is bond, or atom centered or whether a combination of the two occurs.  The comments column refers to points that should be noted in the text about the orbital wave ordering structure.

Figures

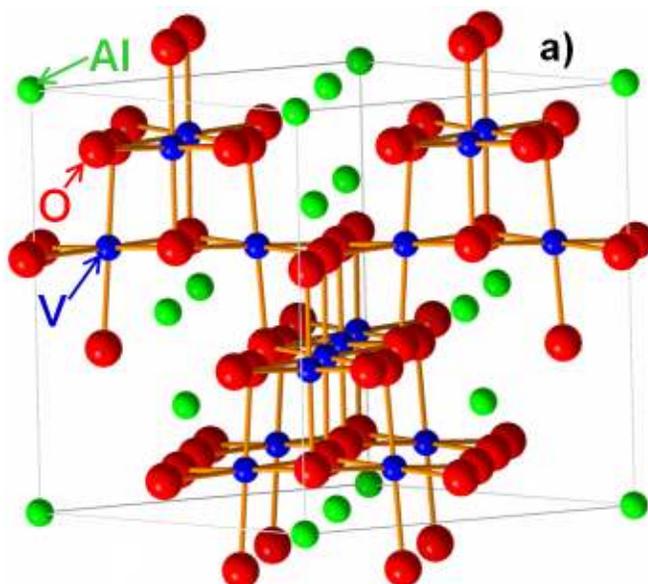

Figure 1a. A traditional view of the unit cell of the high temperature cubic spinel structure of the compound $AlV_2O_4$. Selected O atoms outside of the unit cell have been included to complete the important octahedrally coordinated $VO_6$ moieties. The shared edges of these $VO_6$ octahedra, arranged along the cell edge to edge (e.g. 011) type directions, should be noted. The ligand bonds to the Al atoms, lying in sites of tetrahedral $AlO_4$ coordination, have been omitted.

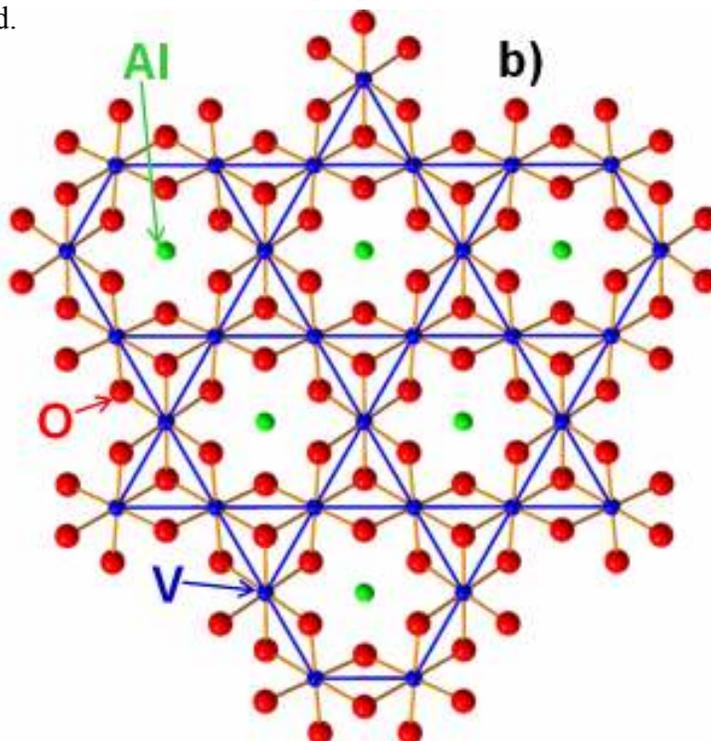

Figure 1b. A view of a single layer of the high temperature cubic spinel structure of the compound $AlV_2O_4$ looking down a body diagonal 111 direction. The Kagome lattice plane of V-atoms is clear from the blue network of lines. The V-sites are all connected across the edges of $VO_6$ octahedra. The ligand bonds to the Al atoms lie directly above and below the hexagonal holes in the Kagome lattice structure.

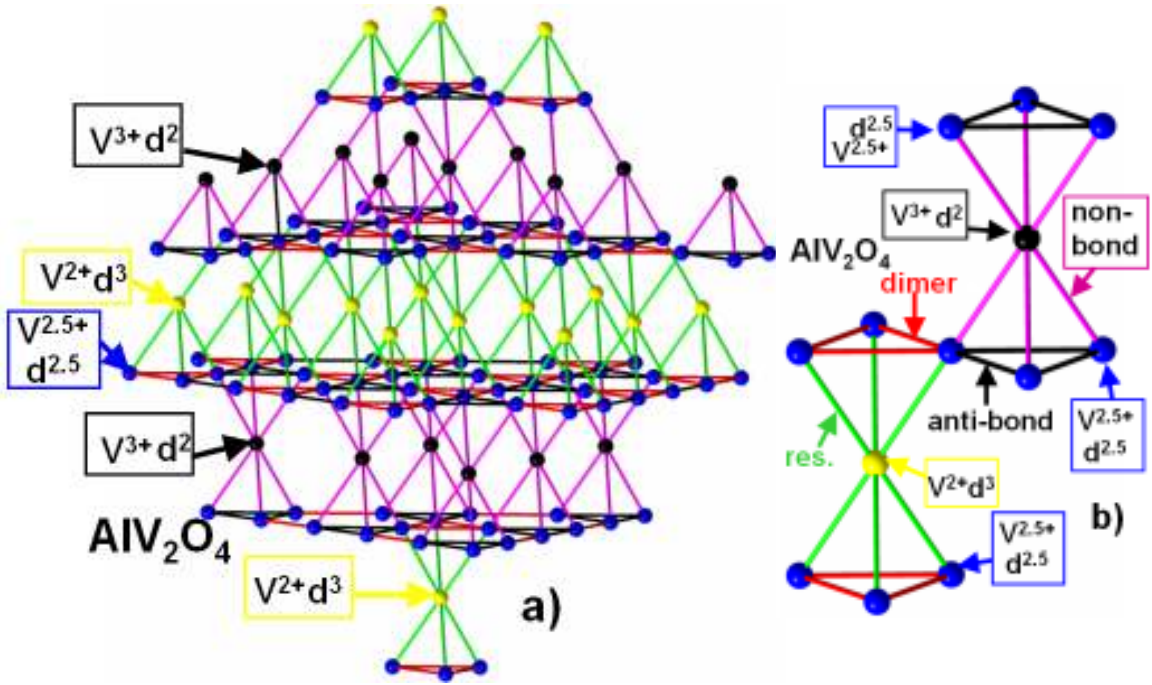

Figure 2a. A wide view of the ordered phase of $AlV_2O_4$ with the cubic 111 direction oriented vertically. Note: 1. the dense Kagome planes of $V^{2.5+}$ atoms (in blue) which contain dimer V-V bonds (red) and anti-bonding (black) V-V pairs; 2. the sparse planes of $V^{2+}$ atoms (in yellow) which bridge the dense planes (from 1) with resonance bonds; and 3. the sparse $V^{3+}$ atom planes (in black) which also bridge the dense planes (from 1) with non-bonding (purple) links.

Figure 2b. An expanded view of the corner sharing tetrahedra linking the $V^{2.5+}$-$V^{2+}$-$V^{2.5+}$-$V^{3+}$-$V^{2.5+}$ sequence of planes. The resonance (res.), dimer, anti-bonding and non-bonding bonds are indicated.

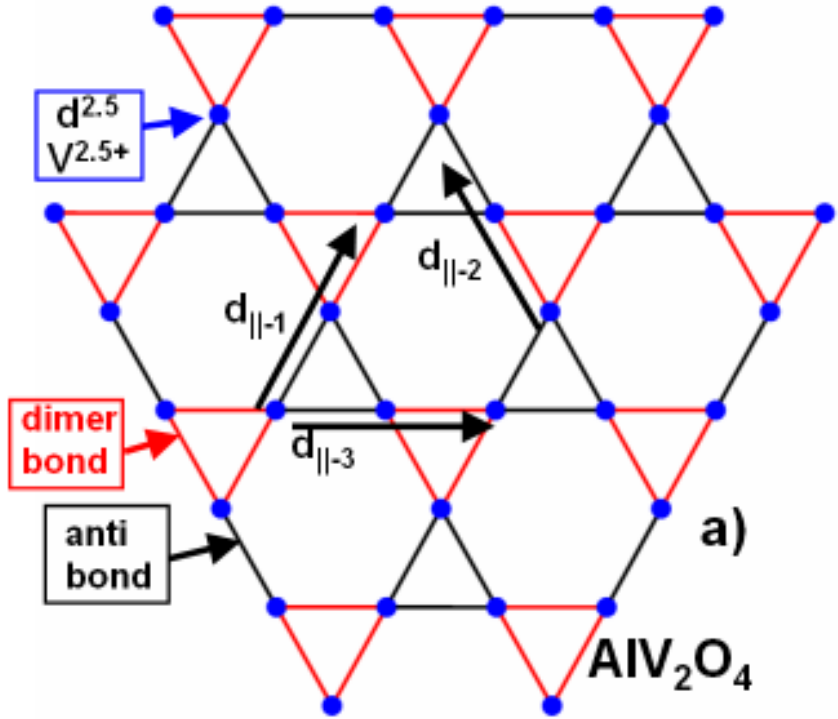

Figure 3a. A view the dense Kagome planes, perpendicular to the [111] direction, of $V^{2.5+}$ atoms (blue) containing dimer V-V bonds (red) and anti-bonding (black) V-V pair links. Three equivalent classes of dimerized chains $d_{\parallel\text{-}1}$; $d_{\parallel\text{-}2}$, and $d_{\parallel\text{-}3}$ with different orientations are indicated

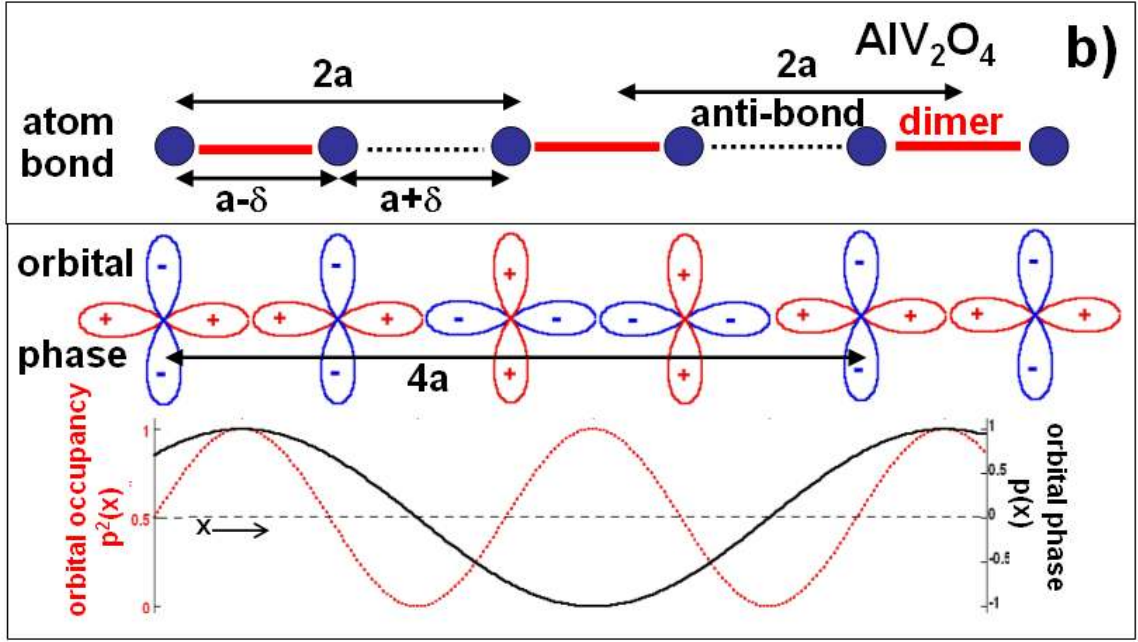

Figure 3b. A schematic of the dimerized $d_{\parallel-1}$; $d_{\parallel-2}$, and $d_{\parallel-3}$ type chains. Top the alternating bond and lattice periodicity of 2a. Bottom: the relative d-orbital phase relationship, p(x), between the sites of the bonding states that are lowered in energy by the dimerization. Note that although the formal repeat distance of 4a for the phase is useful for visualization, the repeat distance for physical observables is 2a. Also shown is the relative d-orbital occupancy relationship, $n_d(d_{\parallel}) = \alpha p^2(x)$ between the sites. Here $\alpha = 2$ to yield the electron count of 1 per site.

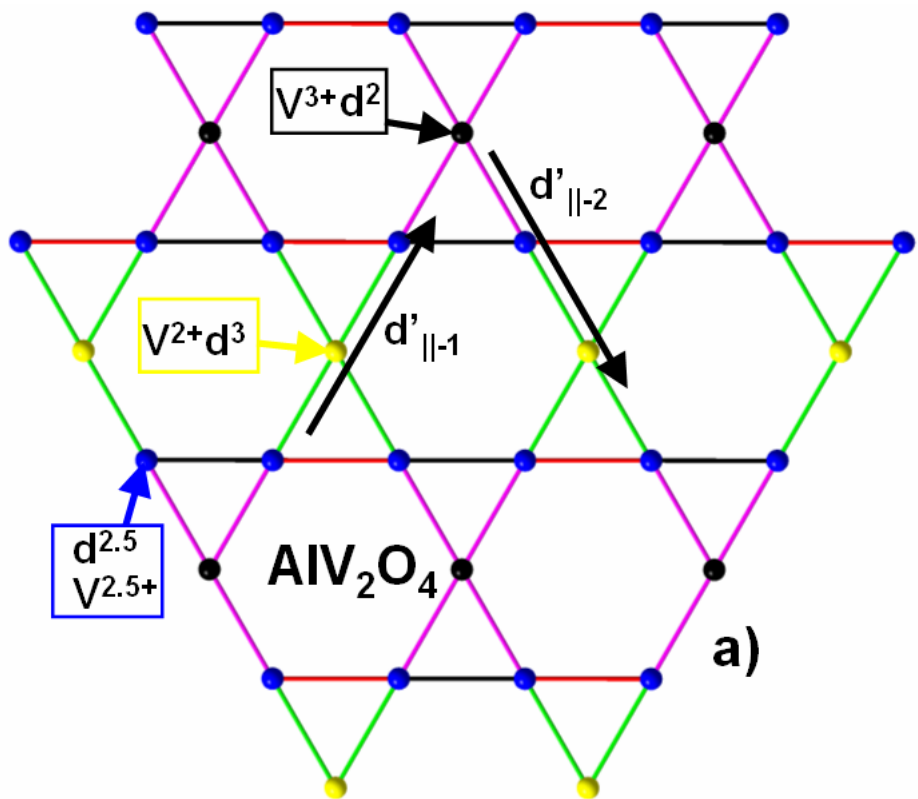

Figure 4a.  A view of the Kagome planes perpendicular to the [11-1] direction.  These planes contain: 1. $V^{2.5+}$ (blue) dimerized atomic chains with alternating V-V pairs dimer bonds (red) and anti-bonding V-V pairs (black) [these chains are the $d_{\|-i}$, i=1,2 and 3 noted in the previous figure]; and 2. tetramerized chains [$d'_{\|-1}$ and $d'_{\|-2}$] with the sequence $V^{3+}$ (black)/non-bond(purple)/$V^{2.5+}$ (blue)/resonance bond (green)/$V^{2+}$ (yellow)/ resonance bond (green)/ $V^{2.5+}$ (blue).

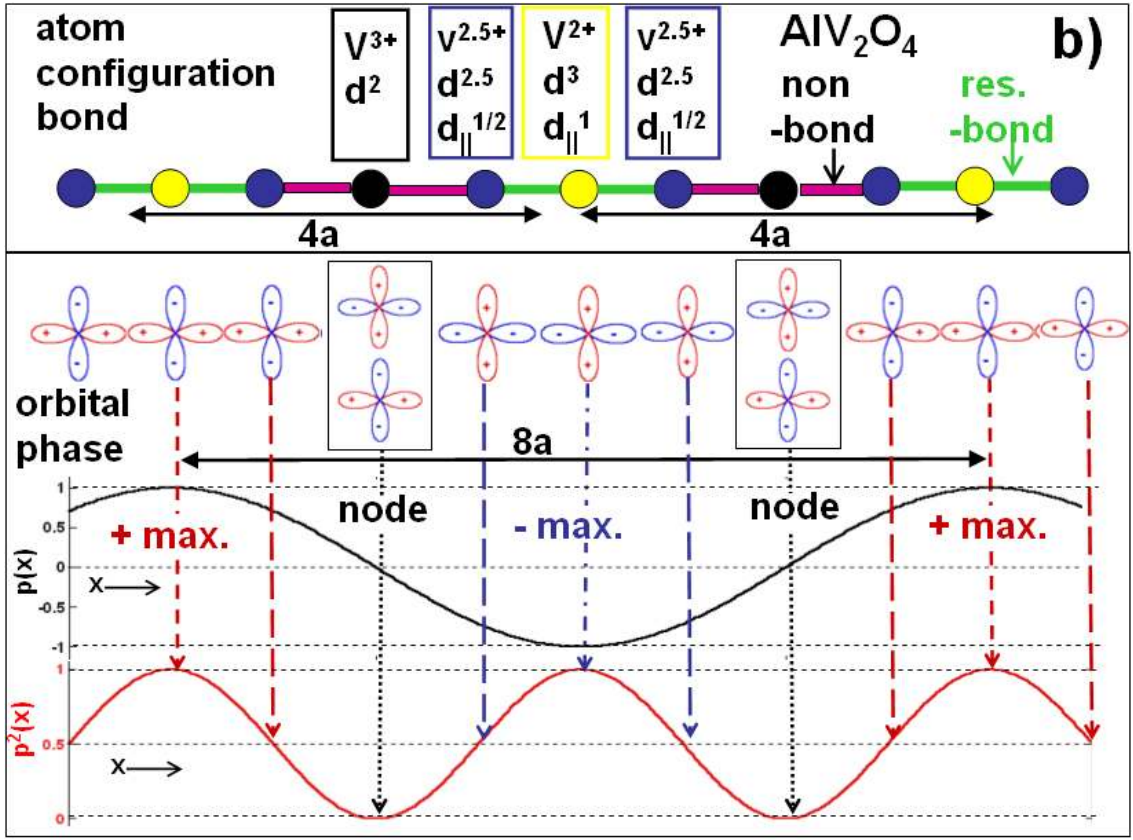

Figure 4b.  A schematic of the tetramerized d'$_{\parallel-1}$; d'$_{\parallel-2}$, and d'$_{\parallel-3}$ type chains.
Top: the 4a periodicity of the tetramerized chain sequence of $V^{3+}$ (black)/non-bond(purple)/$V^{2.5+}$ (blue)/resonance bond (green)/$V^{2+}$ atoms (yellow)/ resonance bond (green)/ $V^{2.5+}$ (blue) is shown.
 Middle: the relative d-orbital phase relationship between the sites in the tetramerized state. Note the formal repeat distance for the phase is 8a, while that of the physical observables is half as long. Note also that at the $V^{3+}$-$d^2$ sites the orbital phase must be frustrated on one side or the other for either the + or − phase choice.  Hence both phases must be equally probable at these sites leaving it in a free decoupled state.  Note that the 8a phase repeat distance was required for this frustrated state at this site.
Bottom: the variation in the relative occupancy $\alpha p^2(x) = n(d_{\parallel})$ of the $d_{\parallel}$ orbital (under the assumption of a quadratic coupling the orbital phase) with its 4a repeat distance.  Here $\alpha=1.0$ to yield the average chain orbital occupancy of ½ per site.

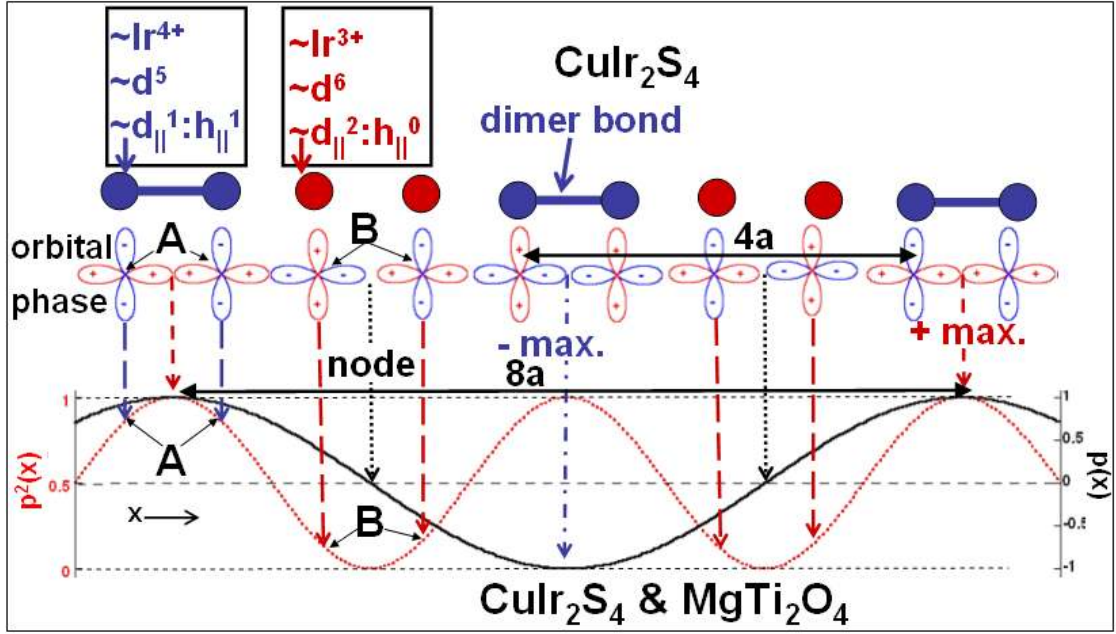

Figure 5. A schematic of the tetramerized Ir-chains in $CuIr_2S_4$. Note that this same schematic can be used for $MgTi_2O_4$ as described in the text.

Top: the 4a periodicity of the tetramerized chain sequence of $Ir^{4+}$-$Ir^{4+}$ dimers [+](blue) and two $Ir^{3+}$ monomers (red).

Middle: the relative d-orbital phase relationship between the sites in the tetramerized state with the repeat distance for the phase being 8a.

Bottom: the variation in the relative orbital phase, p(x), (period 8a) and orbital occupancy $n(d_{\parallel}) = \alpha p^2(x)$ (period 4a) of the $d_{\parallel}$ orbital. Here $\alpha = 1.02$ to yield the average orbital occupancy per chain site of ½.

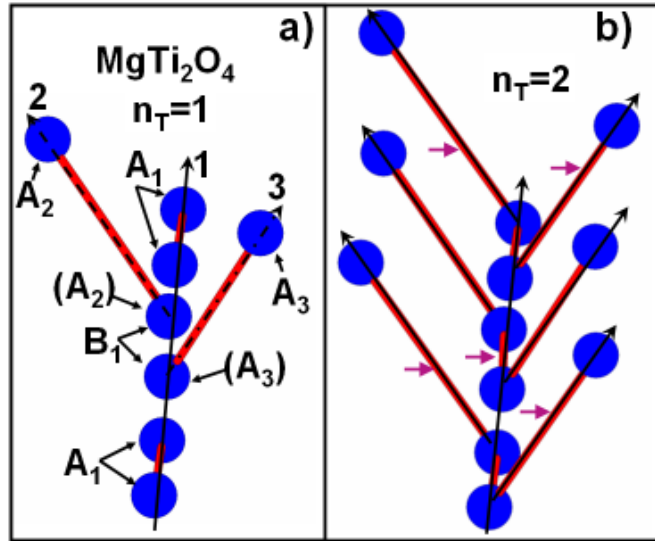

Figure 6a. A schematic of the tetramerized chains i=1, 2, and 3 in $MgTi_2O_4$. Here the $A_i$ sites have dimer bonds between them and the $B_i$ sites have no bond. Note that the B sites are A sites of crossing chains as indicates by the parentheses around $(A_3)$ and $(A_2)$. In the text the A and B site labels in Figure 5 (bottom) will be used to discuss this system. The total electron count $n_T=1$ in this case.

Figure 6b. A generalization of the schematic in Figure 6a in which additional dimer bonds have been added at the positions indicated by violet arrows. This structure now has two dimer bonds per site and a total d-count of $n_T=2$ per site. This structure would be tetragonal as in the previous case.

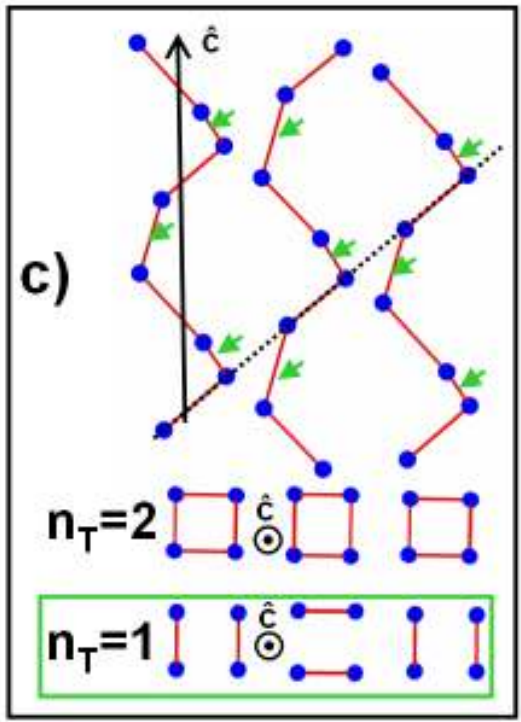

Figure 6c (top). A schematic of the spiraling of the dimer bonds about the tetragonal c-axis (with three spirals shown) for the $n_T=2$ in the case shown in Figure 6b. The dotted line shows one of the dimerized linear chain directions. Note also if the bonds, indicated by green arrows, are dropped one recovers the $n_T=1$ structure of $MgTi_2O_4$.

Figure 6c (middle). A view of the the $n_T=2$ case spirals, shown above, looking down the c-axis.

Figure 6c (bottom). A view of the the $n_T=1$ case (e.g. the structure of $MgTi_2O_4$) spirals, looking down the c-axis.

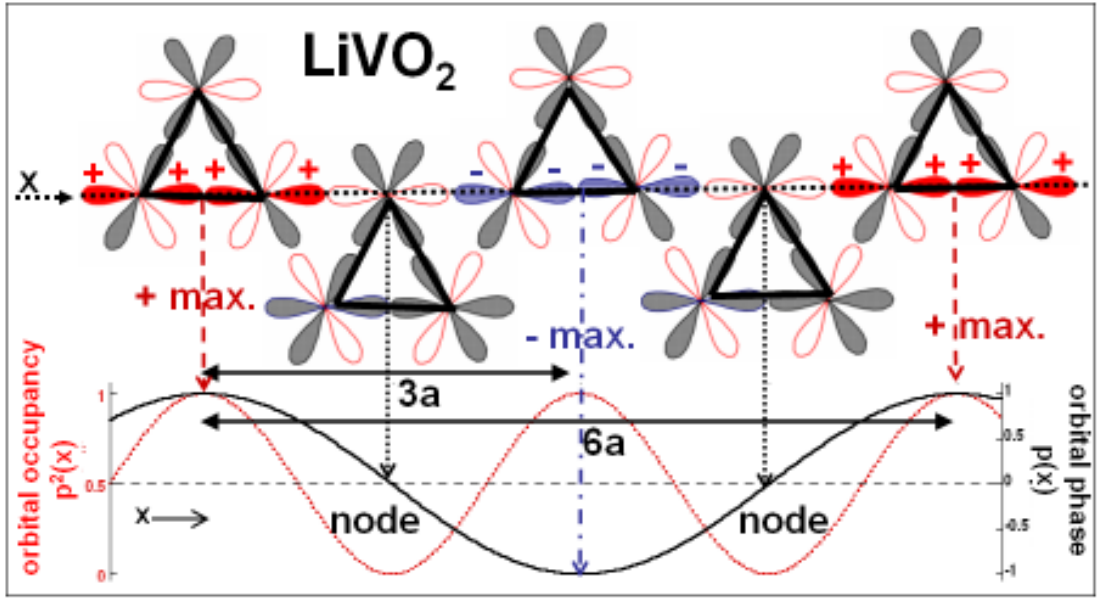

Figure 7. A schematic of the chain type in LiVO$_2$. The filled orbitals are occupied and the empty ones are empty. After Pen et al [24] only the lobes of the d$_{xy}$ type orbitals in the hexagonal V-plane are shown. The solid lines indicate the dimer bonds organized into triangles. The dotted line indicates a representative dimer chain along the x-direction. The relative orbital phase along the typical (horizontal) chain is indicated with the solid red and striped blue being positives and negative respectively. The orbital phase p(x) and orbital occupancy variations [p(x) and $\alpha$p$^2$(x) =n(d$_\parallel$) ] are shown at the bottom of the figure. Here $\alpha$=1.33 would be required to bring the occupancy to 1.0 at the dimer sites.